# Evaluation of a Bundling Technique for Parallel Coordinates

Julian Heinrich, Yuan Luo, Arthur E. Kirkpatrick, Hao Zhang and Daniel Weiskopf

*Abstract*—We describe a technique for bundled curve representations in parallel-coordinates plots and present a controlled user study evaluating their effectiveness. Replacing the traditional $C^0$ polygonal lines by $C^1$ continuous piecewise Bézier curves makes it easier to visually trace data points through each coordinate axis. The resulting Bézier curves can then be bundled to visualize data with given cluster structures. Curve bundles are efficient to compute, provide visual separation between data clusters, reduce visual clutter, and present a clearer overview of the dataset. A controlled user study with 14 participants confirmed the effectiveness of curve bundling for parallel-coordinates visualization: 1) compared to polygonal lines, it is equally capable of revealing correlations between neighboring data attributes; 2) its geometric cues can be effective in displaying cluster information. For some datasets curve bundling allows the color perceptual channel to be applied to other data attributes, while for complex cluster patterns, bundling and color can represent clustering far more clearly than either alone.

## I. Introduction

VISUAL analysis of multidimensional data is required in many applications. As datasets become increasingly large and complex, we need effective ways to display, filter, process, and interpret the information the datasets contain. Many techniques have been proposed for exploratory visualization of multidimensional data, targeted at both generic and specific application domains.[1] One of the main challenges is to provide techniques that scale well with respect to the size of the dataset.

Parallel coordinates are a popular technique for transforming multidimensional data into a 2D image.[2,3] The $m$-dimensional data items are represented as 2D lines crossing $m$ parallel axes, each axis corresponding to one dimension of the original data. This technique has been incorporated into several data visualization and analysis tools, including XLSTAT[4] and GGobi.[5] However, experience has shown several problems with the traditional parallel-coordinates technique. First, the zig-zagging polygonal lines (or polylines, for short) used for data representation are only $C^0$ continuous. They generally lose visual continuation across the parallel-coordinates axes, making it difficult to follow lines that share a common point along an axis—this is known as the cross-over problem.

Second, when two or more data points have the same or similar values for a subset of the attributes, the corresponding polylines may overlap and clutter the visualization. This artifact may occur even for medium-sized datasets with a few thousand points. Finally, clusters and related internal structure of the data are not represented in the geometry of the plot, except for implicit visual clustering based on proximity of polylines at the axes.

Several solutions have been proposed for these problems. The cross-over problem has been mitigated by replacing polylines with smooth curves[7–11] that interpolate the original values at the axes. Cluster perception in parallel coordinates has been facilitated using edge bundling,[12–14] where curves of the same cluster are grouped geometrically. In contrast to the traditional color-coding of clusters, the resulting curve bundles also reduce visual clutter by freeing up plot space to provide an overview of the data.

We propose a variant combining the benefits of both the above approaches. We use a curve model based on piecewise cubic Bézier curves that supports bundling at the cluster centroids. Given a clustering of the data, our method allows fast construction of the curve bundles while guaranteeing good visual continuation of the lines. In addition to parametrized smoothness of lines,[11] it also allows for tuning the bundling tightness, providing a range of representations from pure polylines to tightly-bundled curve plots, enabling the user to obtain different views of the data.

We also conducted a controlled user study to compare the effectiveness of polylines and our curve bundling technique with respect to cluster perception and correlation judgment. While variants of polylines and curves have been evaluated[11,15] no prior study evaluated the joint effect of these two features on the perception of clusters and correlations. See Table I for a summary of the evaluations.

The study showed that curve bundling maintains the users' ability to recognize correlation between data attributes, a traditional strength of parallel coordinates. Furthermore, for revealing clusters to the user, curve bundling is at least on par with color coding, the traditional way of representing clusters.

TABLE I
EVALUATIONS OF PARALLEL COORDINATES.

|  | Correlation | Cluster Identification |
|---|---|---|
| Polylines | Li et al.[15] | Holten et al.[11] |
| Curves | This paper | Holten et al.[11] |
| Bundling | This paper | This paper |

Arthur E. Kirkpatrick, Yuan Luo, and Hao Zhang are with the School of Computing Science, Simon Fraser University, Canada. E-mail: {ted,yuanl,haoz}@cs.sfu.ca.

Julian Heinrich, Daniel Weiskopf are with VISUS (Visualization Research Center), Universität Stuttgart, Allmandring. 19, 70569 Stuttgart, Germany. E-mail: {weiskopf,julian.heinrich}@visus.uni-stuttgart.de.

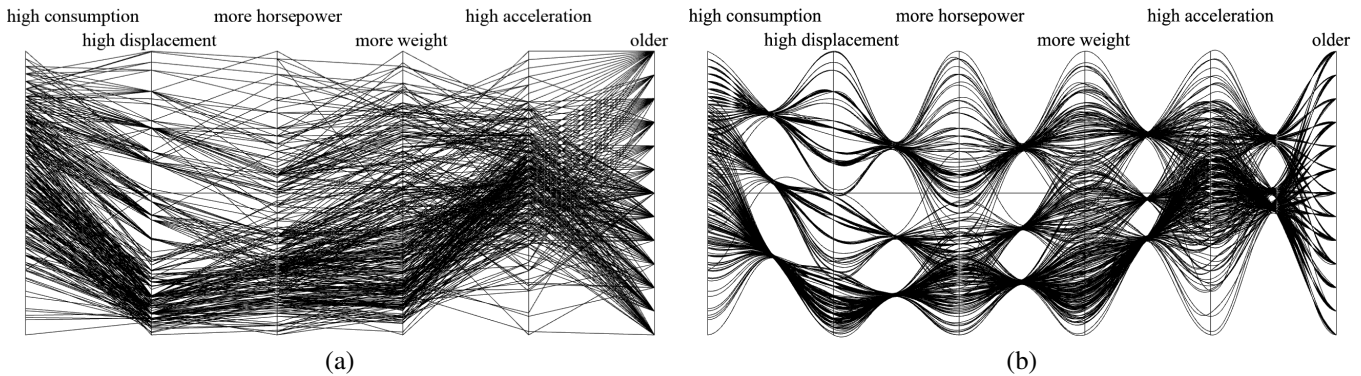

Fig. 1. The cars[6] data displayed as (a) polyline and (b) and bundled plots. Data are clustered by number of engine cylinders (4, 6, or 8). In the bundled plot, bundling was $\beta = 0.95$ and cluster centroids were plotted at their projected values on the bundle axis. The adjectives above each value axis indicate the interpretation of values closer to the axis top.

Figure 1 compares the polyline version of parallel coordinates with a version using our technique of bundled curves.

The paper begins with a summary of prior work on polyline parallel coordinates and variants using curves. Section III describes the mathematics of our bundled curves, while Section IV presents the design of the user study and its results. This is followed by an extended example in Section V, showing how bundled curves facilitate detailed analysis the substructure of clusters in a seven-dimensional data set. This description is supplemented by a video, available at the supporting material site. The paper concludes with a summary of the argument and proposals for future work.

Throughout the paper, we have used example data sets that are of medium size, ranging from 40–400 data points and three to nine clusters. We believe that this size allows us to compare the polyline and bundled representations most directly. Data sets with larger numbers of points or clusters will typically require a sophisticated analysis using multiple representations and extensive interaction by the analyst, making it difficult to assess the specific contribution of line or curve representation. The more modest data sets presented in this paper allow focused comparison and attribution of results to the representation.

## II. Related Work

In parallel-coordinates visualization, points in $m$-dimensional space are represented as lines crossing $m$ parallel axes in 2D, so there is no inherent limit on dimensionality. The process of discovering multivariate relations in a dataset is transformed to a 2D pattern recognition problem. Parallel coordinates were introduced by Inselberg,[2,3,16] and extended by Wegman.[17]

Traditional parallel coordinates suffer from several problems, especially for large datasets. One issue is the potentially heavy over-plotting of lines, resulting in visual clutter. A proposed remedy is to replace fully opaque, rasterized lines by a density representation of the plotted lines.[18,19] This idea has been adopted for frequency plots,[20] gray-scale mappings in density plots,[21] and high-precision textures in combination with transfer functions.[22] For continuous data, line density can also be computed analytically using an appropriate reconstruction kernel.[23]

The cross-over problem for polylines arises when two or more lines share common points on an axis. Several authors have solved this by using smooth curves. Theisel[7] proposes a cubic B-splines model, while Graham and Kennedy[8] choose a quadratic or cubic curve for a particular section depending on the shape formed by that section and the two adjacent sections. Moustafa and Wegman[9] build smooth curves by replacing the piecewise linear interpolation of polylines by interpolation via higher-order sinusoidal functions. Holten[11] adds a parameter to the spline-based models[8,10] to control the amount of smoothing. All these techniques guarantee curve smoothness, alleviating the cross-over problem by giving different trajectories to points that intersect at an axis. This allows the analyst to reliably connect the curves on either side of the axis.

Visual clutter can also be reduced by preprocessing the data with a clustering algorithm.[24] The clusters can then be displayed by extensions of parallel coordinates.[19,25,26] Whereas early clustering work focused on reducing the amount of displayed data by displaying only markers of entire clusters, recent work has instead focused on displaying all the data and revealing details of the internal structure of clusters. Johansson et al.[22] combine specific transfer functions for density plots with feature animation, showing both an overview of the data and the inner structure of its clusters. Novotny and Hauser[27] extend such cluster-based parallel-coordinates visualization to additionally display outliers and trends. Zhou[28] detects clusters by splatting lines and applying a Gaussian weight to proximate lines.

Two recent publications[13,14] enhance parallel-coordinates plots following the same perceptual motivation of geometric proximity as our method. Zhou et al.[13] deform traditional polylines by applying attracting and repelling forces. By construction, their method is based on proximity between the initial polylines and, thus, achieves an implicit, yet fixed type of clustering. Their method emphasizes the proximity of the polylines, rather than showing externally provided clusters. Moreover, their visual clustering is based on a piecewise model: the vicinity of polylines between two neighboring data axes (or dimensions) of the parallel-coordinates plot governs the visual clustering between those two data dimensions; other pairs of neighboring data dimensions are clustered independently. Therefore, high-dimensional data is not clustered on a per-data-point level, but based on pairs of data dimensions. The resulting visual clustering is thus sensitive to the order of data dimensions in the parallel-coordinates plot.

Holten introduced edge bundling of tree layouts.[12] McDonnell and Mueller[14] built on this idea, developing a geometric, spline-based approach to computing visual bundling that is similar to ours. However, McDonnell and Mueller's technique has a different objective: it targets illustrative parallel-coordinates plots, using visual simplification and non-photorealistic rendering techniques such as silhouette lines, halos, and shadows. Therefore, details of the internal structure of data points within clusters are not a focus of their research. Moreover, cluster membership information is still based on color coding, whereas our approach provides a complementary, geometry-based visualization of clusters.

Our method improves upon the proximity-based parallel-coordinates techniques of McDonnell and Mueller[14] and Zhou et al.[13] in the following ways. First, we make better use of the available screen space by re-distributing visually clustered curves in a uniform way. Therefore, there is much less overlap in the important parts of the plots—in the regions between two data axes, where users identify correlation of data points. In addition, overdraw and cluttering issues are reduced by this redistribution. Second, we guarantee $C^1$ continuity of the curved lines across data axes, addressing the cross-over problem.

There have been few previous papers providing user studies on parallel coordinates. Jing[15] compares polyline parallel coordinates and scatterplots. Lanzenberger et al.[29] investigate the effectiveness of stardinates and parallel-coordinates plots applied to an example data set with psychotherapeutic data. Henley et al.[30] evaluate scatterplots and parallel coordinates for the task of comparing genomic sequences. Ten Caat and Roerdink[31] study a tiled parallel-coordinates technique for visualizing time-varying multichannel EEG data. Johansson et al.[32] investigate the amount of noise that may be present in parallel coordinates such that patterns can still be received. Holten and van Wijk[11] evaluate cluster identification performance for curved and animated parallel coordinates, among others. While Li and van Wijk[15] examined the visualization of correlation for linear parallel coordinates and Holten and van Wijk[11] the visualization of clusters, our user study aims at evaluating the impact of bundling and curves to the judgment of correlation and the detection of clusters.

Finally, there seems to be no literature concerning the evaluation of bundling at all.

## III. CURVE MODEL AND CURVE BUNDLING

In this section, we extend previous curve models[7,11] to support bundling curves with the same cluster membership. The extension is required because previous bundling techniques for parallel coordinates either use a different curve model[14] or a different clustering model.[13]

The bundled curve model uses curve geometry to increase the visibility of structure in the data across multiple axes, reveal structure within clusters, and alleviate the cross-over problem. At the same time, curve control points and tangents are chosen to achieve good approximation to the original polylines. The construction is designed to maintain the desirable characteristics of polyline plots, in particular, their ability to reveal correlations between variables. Finally, the parameters of the model support exploration of the data by adjusting its representation.

We use terminology and background knowledge from geometric modeling, especially Bézier curves and the concept of parametric ($C$) continuity.[33]

### A. Overview

Given an input of $N$-dimensional data points with specified cluster membership, denote the parallel-coordinates axes by $X_1, X_2, \ldots, X_N$. We refer to these as the *value axes* to distinguish them from the additional axes created for curve modeling and bundling. Assume (without loss of generality) that the value axes are uniformly distributed across the width of the plot and separated by unit intervals. Given a data point $(P_1, P_2, \ldots, P_N)$, its corresponding polyline is replaced by a piecewise cubic Bézier curve with the following properties:

- The curve *interpolates* $P_1, P_2, \ldots, P_N$ at the value axes.
- The curve is $C^1$ continuous throughout.
- Curves corresponding to data points that belong to the same cluster are bundled between adjacent value axes. This is accomplished by inserting a *bundle axis* midway between the value axes and by appropriately positioning the Bézier control points.
    - To support curve bundling, curves within a given cluster are adjusted by moving their control point on the bundle axis toward the value of their cluster centroid on the bundle axis.
    - The cluster centroid is the projection of the N-dimensional centroid on the plane defined by the respective adjacent value axes, intersected with the bundle axis.
    - The use of curves also allows us to take better advantage of the entire plot area. Specifically, the cluster centroids can be arbitrarily distributed along the bundle axis to alleviate the line clutter problem.
- Two parameters, $\alpha$ and $\beta$, adjust the shape of the Bézier curves. The parameter $\alpha$ controls the extent the curve approximates the original piecewise linear polyline, while retaining $C^1$ continuity; we call $\alpha$ the *smoothness scale*. The parameter $\beta$ dictates the *bundling strength*, how tightly the curves within a cluster are pulled together. It is worth noting that polyline-based parallel-coordinates plots are simply a special case of our curve model, when we set $\alpha = \beta = 0$. Other curve models[11] are obtained by setting $\beta = 0$.

### B. Curve continuity and smoothness scale

Consider two adjacent value axes $X_i$ and $X_{i+1}$ with points $P_i$ and $P_{i+1}$ on them, respectively. Let the intersection of the line segment $P_iP_{i+1}$ with the bundle axis $V_i$ (halfway between $X_i$ and $X_{i+1}$) be $Q_i$ (Figure 2). We shall convert the straight line segment $P_iP_{i+1}$ into two cubic Bézier curve segments: $b_{i,1}$ between $X_i$ and $V_i$ and $b_{i,2}$ between $V_i$ and $X_{i+1}$. Due to curve bundling, we may move $Q_i$ to $Q'_i$ along $V_i$ as a result of attraction by a cluster centroid $C_i$; this is explained in Section III-C. The new point $Q'_i$ will serve as a common control point for the two Bézier curve pieces $b_{i,1}$ and $b_{i,2}$, which are constructed using Holten and van Wijk's method[11] by setting $p_0 = P_i$ and $p = Q'_i$.

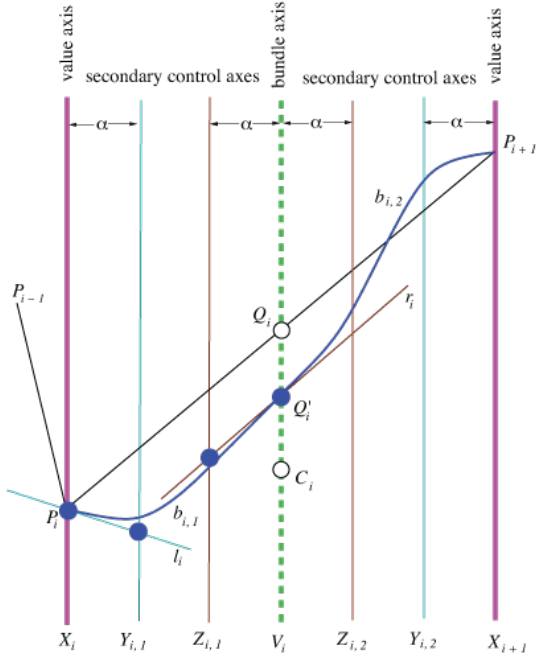

Fig. 2. Construction of Bézier curve pieces between two adjacent value axes $X_i$ and $X_{i+1}$. We insert a bundle axis $V_i$ midway between $X_i$ and $X_{i+1}$, as well as secondary control axes, $Y_{i,1}, Z_{i,1}, Z_{i,2}, Y_{i,2}$, placed at a distance $\alpha$ away from their corresponding value or bundle axis. Adjacent Bézier pieces share the same tangent line and intersections between the tangent lines, while the value, bundle, or secondary control axes define the control points for the Bézier curves. The control points for $b_{i,1}$ are shown as blue dots. The control points for $b_{i,2}$ would be at corresponding locations on the right. The point $C_i$ is a cluster centroid which attracts the constructed Bézier curves in curve bundling. As a result, while the original polyline passes through $Q_i$, the curve pieces now pass through by $Q'_i$.

### C. Curve bundling and bundling strength

The model can be extended to represent cluster membership using bundling. Consider the set of polylines belonging to a particular cluster $\mathscr{C}$. For each such polyline, we record its intersections with all the bundle axes. For a particular bundle axis $V_i$, we can use the centroid $C_i$ of the intersection points corresponding to the polylines in cluster $\mathscr{C}$ as the common control point shared by all the Bézier pieces (those which replace polylines in cluster $\mathscr{C}$) adjacent to $V_i$. This will force all the Bézier curves in the cluster to pass through the common point $C_i$.

Strict curve bundling may hinder the viewer's ability to distinguish a positive correlation from a negative one. This is illustrated in Figure 3, where the first two columns show the polyline plots (Figure 3a) and strictly bundled plots (Figure 3b) on two variables with strong positive correlation, no correlation, and strong negative correlation. It is still possible to detect the minor differences in the bundled plots, but not as easily as for the polyline plots.

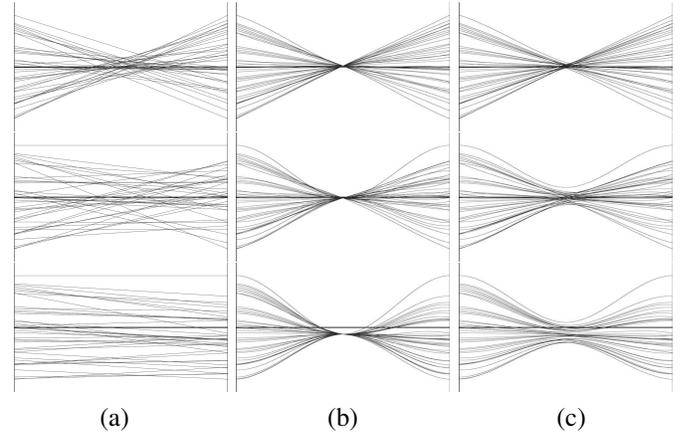

Fig. 3. Effect of bundling strength $\beta$ on representing correlations. Column (a) shows polyline plots of two variables with strong negative correlation, no correlation, and strong positive correlation, respectively; Column (b) shows the corresponding bundled curve plots with strict bundling, $\beta = 1$; Column (c) shows bundled curve plots with $\beta = 0.8$.

As a remedy, we introduce the bundling strength parameter $\beta$. As shown in Figure 2, let the intersection between a straight line segment and the bundle axis $V_i$ be $Q_i$. The bundling strength controls the extent to which $Q_i$ will be pulled or attracted toward the centroid $C_i$ defined above. Currently, we apply a linear weighting scheme: the new $Q'_i$ will be given by

$$Q'_i = (1-\beta)Q_i + \beta C_i,$$

where $0 \leq \beta \leq 1$. The choice $\beta = 1$ corresponds to strict curve bundling, whereas $\beta = 0$ disables bundling altogether. Any reasonable nonlinear weighting scheme, e.g., Gaussian weights,

$$Q'_i = (1-\eta)Q_i + \eta C_i, \quad \text{where } \eta = \exp(-||Q_i - C_i||^2/\sigma^2),$$

for a suitably chosen width $\sigma$, is possible. The same range of weighting scheme can be used for the centroid $C_i$.

In our current implementation, the same $\beta$ is applied to all instances of curve bundling. Obviously, one can gain more fine-grained control by tuning $\beta$ individually for specific curves, bundle axes, or clusters; this would be at the expense of introducing a more complex control interface.

By adjusting the bundling strength properly, we can detect correlations in bundled curve plots just as easily as in polyline plots, even in the presence of clear curve bundling. As shown in Figure 3c, with $\beta = 0.8$ more negatively correlated data form denser, narrower crossing bands. A quantitative, empirical comparison of the effectiveness of bundled curve and polyline for discovering variable correlations is given in the next section.

*D. Redistribution of cluster centroids*

To take advantage of the entire available plot area and further alleviate the line cluttering problem, the cluster centroids can be sorted along each bundle axis and then redistributed *uniformly* across the length of the bundle axis in an *order-preserving* manner, where order is separately defined by the centroid for every adjacent of axes. Figure 5 shows an example of redistributed centroids. An immediate extension to uniform distribution is to take into account the sizes of the clusters and adjust the positioning of the cluster centroids accordingly.

A given analysis may or may not benefit from distributing the centroids. Cluttered plots will have their clutter reduced through distribution. On the other hand, for plots with greater separation between clusters, plotting the centroids at their projected values can provide useful information for the analyst. We have used both methods in this paper: The user studies and most example plots used distributed clusters, while the example application in Section V does not use distributed clusters.

## IV. USER STUDY

To compare the effectiveness of polylines and bundled curves, we performed a user study. Observers were asked to estimate (a) correlations and (b) the number of clusters, in datasets represented using polylines and bundled curves. We expected that bundled curves would support correlation estimation at least as well as polylines do, and that bundled curves would support superior estimation of the number of clusters.

*A. Overview*

In designing the experiment, we were subject to the constraint that we needed to estimate performance by analysts skilled both in an underlying domain and at interpreting parallel coordinates a given type, polyline or bundled curves. Such users are not merely difficult to find, for the case of bundled curves they do not yet exist. We addressed this constraint with an approach often used for visualization user studies. Specifically we:

1) Recruited participants who had little to no experience with either form of parallel coordinates. We gave the participants a short tutorial on strategies for estimating correlations in parallel plots of each style. This created a pool of participants equally skilled at reading both styles, somewhere between novice and intermediate skill.
2) Used data sets generated solely according to specified probability distributions, with no underlying semantics. This ensured that no participant would be able to apply domain knowledge to interpret the plots.

We used accuracy as the sole dependent measure and did not record time. We argue that this untimed task matches the context in which data analysts typically use parallel coordinates, taking enough time to consider their data in depth. This choice emphasized that participants take as long as necessary to make their best estimate. It also minimized fatigue by allowing participants to rest whenever they wished, without regard for their score. This choice also eliminated the potential confound of different participants adopting different speed-accuracy tradeoffs, because accuracy was uniformly emphasized.

Given the limited experience of the participants with the two styles of plot, we do not believe that timing data would provide any useful comparison between the styles. Comparative timing data would only be informative with testers who were well-experienced with the methods, working on data sets for which they had domain expertise.

The curve styles were compared for two tasks, estimating correlation and estimating number of clusters. The tasks were performed in a fixed order for every participant, with participants estimating correlation first. This design permits more direct interpretation of the results because all participants performed each task with a fixed level of prior experience. In particular, their experience reading plots in the correlation task would carry over to enhance their performance in the cluster estimation task.

By contrast, a design that counterbalanced task order would have split participants' prior experience, increasing the variance and making the results harder to interpret. Given that a counterbalanced design would only protect against the case that doing the correlation estimate first would *differentially* advantage one curve style, a prospect we consider highly unlikely, we chose a fixed task order for its more straightforward interpretation.

*B. Design*

The study design was single-factor, two-level, and within-subjects. Observers viewed two data series. The first was always the correlation estimation series, the second the cluster estimation series. Within each series, line style was a blocked factor, with all trials performed first in one line style, then the other. Order of the two line styles was counterbalanced across participants, with participants randomly assigned to the order. Dependent measures, computed separately for each series, were the Pearson correlation $r$ between the actual dataset correlation and the correlation estimated by participants, and the Fleiss $\kappa$ measure of agreement amongst participants.

Before running the full study, we ran a pilot study with five participants to determine the best values of bundling strength $\beta$ and smoothness scale $\alpha$. The values $\beta = 0.8$ and $\alpha = 1/6$ achieved the best balance of correlation detection and cluster visualization. These values were used for the bundled plots in both series of trials.

*1) Participants:* A convenience sample of 14 participants (9 men, 5 women, ages 23–37) was recruited from graduate students in computing science and engineering science at Simon Fraser University. Of these 14, 2 had previously used polyline parallel coordinates, 8 had experience with some form of information visualization but had never used parallel coordinates, and 4 had never used any visualization software. Volunteers were paid CDN$ 20.

*2) Procedure for the session:* Participants first answered a brief series of questions assessing their level of experience with information visualization and computers in general. They were next tested for any color deficiencies using a Web-based test.[34] All 14 participants had acceptable color vision. They next read a tutorial on the basic principles of parallel coordinates and their instantiation in polylines and bundled curves. The tutorial defined correlation and gave examples of positive and negative correlation using both line types.

Participants then began the first series of trials, in which they estimated correlations. Immediately after completing that series, participants began the second series, in which they estimated the number of clusters in plots. After completing the second series, they indicated which line style they preferred and answered a short list of open-ended questions about their experience during the study.

Participants were allowed to take as long as they wished on each trial. Total time to complete the session varied widely, from 50 to 110 minutes. Most participants completed the study in less than 90 minutes.

### C. First series: estimating correlations

In the correlation estimation trials, participants viewed a series of datasets in 2D parallel coordinates (i.e., with two data dimensions and two main parallel-coordinates axes), plotted in either polylines or bundled curves. For each plot, the user was asked to categorize the correlation as "strong negative correlation", "negative correlation", "no correlation", "positive correlation", or "strong positive correlation".

*1) Procedure:* Before starting each line style, participants read a short tutorial on estimating correlations in that style. For polylines, the tutorial suggested looking for whether the lines crossed or not, the distribution of line crossings (whether only in the middle or distributed throughout the range), and the overall shape of the plot. For bundled curves, the tutorial suggested looking at the width of the middle band and the overall shape of the plot. For each style, the tutorial presented example plots of all seven degrees of correlation. To map the seven values of actual correlation to the five categories of user response, the tutorial recommended reporting $z$ values of $-1.0$ and $-0.5$ as both "negatively correlated", and similarly for $+0.5$ and $+1.0$.

Participants began each line style with a training session. The training session presented one plot of each correlation level in the given style. Participants estimated the correlation and were then told which answers would have been appropriate for the dataset. Since there were seven levels of correlation but only five levels of user response, two possible answers were suggested for every example. After estimating all seven practice correlations, a page was displayed reminding participants of the strategies for estimating correlation for this plot style. Users pressed a button to start the first experimental trial.

Experimental trials had the same interface as the practice trials, but provided no feedback about the actual correlation. When the participant was satisfied with their estimate for the current trial, they pressed a button to start the next.

*2) Trial data:* Three groups of seven datasets were generated, each with $n = 40$ data pairs. The pairs were generated from normally distributed random series $x$ and $y$, selected to ensure that each set of 40 pairs had the given correlation coefficient. Each group of datasets had exactly one set for each level $z = -1.5, -1.0, -0.5, 0.0, +0.5, +1.0, +1.5$, where $z$ is the Fisher transform of the correlation. These were the same levels of correlation used in prior work.[15] One group of datasets was always used for the training phase. The remaining 14 datasets were each used twice, once for each line style. Within each line style, order of datasets varied randomly for each participant.

The bundled curve representation required two additional parameters for each data point: the directions of the line leaving each axis. These are not required for polyline plots, where the direction of the line leaving an axis is independent of the direction the line entered that axis from the other side. However, the exit direction of bundled curves is affected by the direction of the line entering from the other side, due to the $C^1$ continuity requirement. Pilot tests showed that if all curves entered the axes at a constant horizontal direction, observers used the consistent bending of curves at the axes as a cue to estimate correlation. Since this cue would not occur in actual use of bundled curves, which would in fact enter their axes at varying angles, the direction at which each curve entered each axis was randomly perturbed. This random perturbation likely made correlation detection slightly more difficult for bundled curves than it would be in practice, where entry to the axes would vary but not be random.

Figure 4 illustrates example datasets with all seven different correlation coefficients used for the experimental trials. The top row shows polyline plots and the bottom row shows bundled curve plots.

### D. Second series: estimating clusters

In the cluster estimation trials, participants viewed a series of clustered datasets in parallel coordinates ranging from two to six dimensions, plotted in either polylines or bundled curves. Clustering was indicated by color (for polyline plots) and bundling (for bundled curve plots). Color Brewer[35] was used to define effective color maps for the polyline plots. For each plot, the user was asked to estimate the number of clusters.

*1) Procedure:* Before starting the series, participants read a description of how clusters are represented in both line styles. They then began working with either polyline plots or bundled curve plots, depending upon which order had been assigned. They practiced estimating the number of clusters in three trial plots, with five, three, and eight clusters. Figure 5 shows typical examples of such plots. After each training trial, the correct number of clusters was reported. After the three training trials, a page redisplayed the three datasets and the number of clusters in each. Users pressed a button to start the first experimental trial.

Experimental trials had the same interface as the training trials, but provided no feedback about the actual number of clusters. After entering their estimate for the clusters, participants pressed a button to move on to the next trial. Once they had completed a series in one line style, they did the training and experimental trials for the next style.

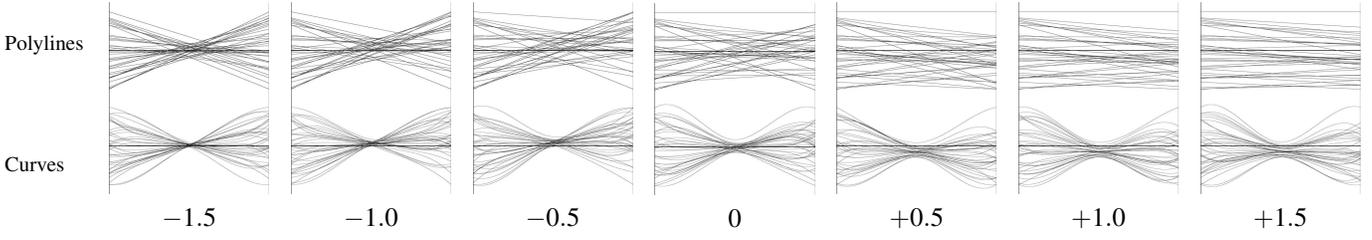

Fig. 4. One of the three sets of plots used in the correlation estimation series. The correlation, specified in Fisher $z$, is shown below each plot.

TABLE II
DATASETS FOR THE CLUSTER ESTIMATION SERIES ($d$ IS THE NUMBER OF DIMENSIONS, $n$ IS THE NUMBER OF DATA POINTS)

| Name | $d$ | $n$ | Source |
|---|---|---|---|
| iris | 4 | 150 | Botany |
| netperf | 6 | 179 | Computer Science |
| htong | 4 | 365 | Earth Science |
| g40 | 2 | 40 | Synthetic |
| g160 | 3 | 160 | Synthetic |
| g200 | 5 | 200 | Synthetic |

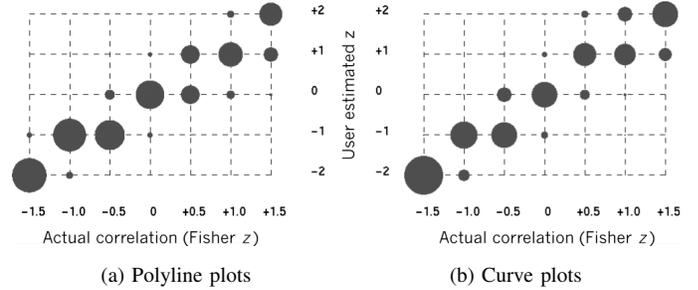

(a) Polyline plots  (b) Curve plots

Fig. 6. Distribution of responses for the estimated correlations of 2D parallel coordinates for polyline plots and bundled curve plots. Circle radius represents the frequency with which participants estimated a correlation strength for each actual correlation.

*2) Trial data:* Trial datasets were created from three real-world and three synthetic datasets (Table II). The real-world datasets are popular test datasets, taken from the Xmdv Web page.[36] The synthetic datasets were generated by sampling normally distributed series, selected to ensure the required correlation across each dimension. Each of the 6 datasets was then clustered by the $k$-means technique into $k = 3, 5, 7,$ and $9$ clusters. This series of 24 datasets was plotted using both line styles. Within each series, the order of trials varied randomly for each user.

### E. Results

Figure 6 shows the distribution of participants' responses for the correlation estimation series. There was a strong linear correlation between participants' estimates and actual correlation for polylines and bundled curves (both $r = 0.90$). Considering the estimates for positive and negative correlations separately, estimates for negative correlations were stronger ($r = 0.75$ for polylines, $r = 0.79$ for bundled curves, difference of the equivalent $z$-scores $\Delta z = 0.10$) than for positive correlations ($r = 0.55$ for polylines, $r = 0.39$ for bundled curves, $\Delta z = -0.21$). Agreement amongst participants was moderate ($\kappa = 0.43$ for polylines, $\kappa = 0.41$ for bundled curves). The results for polylines are comparable to those of Li and van Wijk.[15]

For all comparisons the one-sided width of a 95% confidence interval, which is entirely determined by the sample size of 14, is $\Delta z_{95\%} = 0.59$. All the $\Delta z$ scores presented above were substantially within this bound, indicating that none of the differences was statistically significant.

Figure 7 shows the distribution of participants' responses for the cluster estimation series. The overall correlation is strong for both line styles ($r = 0.92$ for polylines, $r = 0.96$ for bundled curves, $\Delta z = 0.36$). The correlations were much stronger for datasets with three or five clusters ($r = 0.98$ for both line styles) than those with seven or nine clusters ($r = 0.41$ for polylines, $r = 0.68$ for bundled curves, $\Delta z = 0.39$). As with the correlation estimation series, all $\Delta z$ values were substantially below 0.59, indicating that none of the differences was statistically significant.

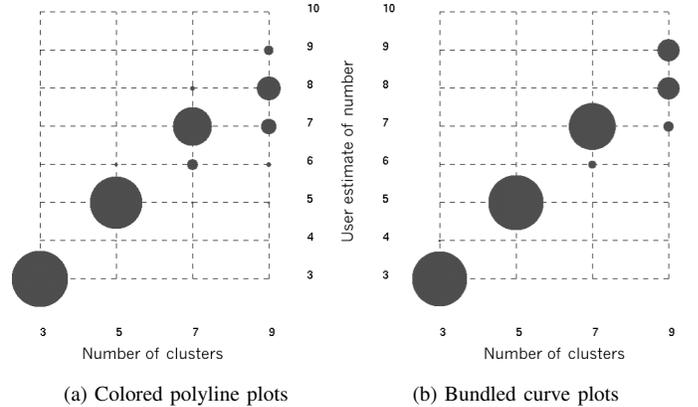

(a) Colored polyline plots  (b) Bundled curve plots

Fig. 7. Distribution of responses for the estimated number of clusters in parallel-coordinates plots in two line styles. Circle radius represents the frequency with which participants estimated a dataset to have a given number of clusters.

Agreement amongst participants for cluster estimation was slightly higher for bundled curves ($\kappa = 0.65$) than for polylines ($\kappa = 0.56$). Each line style had higher agreement than their corresponding levels for correlation estimation.

### F. Discussion

The results for the two series of plots demonstrate important strengths of the bundled curve representation. The correlation estimation series demonstrates that correlation is as readily recognizable when parallel coordinates are rendered in bundled curves as when rendered in polylines. This result is not obvious. Polyline plots provide a clear focal point for estimating correlations: the width of the center region is an excellent indicator of correlation, with strong negative correlations producing a narrow center region and strong positive correlations producing a wide center region. In contrast, bundled curve plots by definition draw the curves into one or more narrow center regions. The width of those regions is only mildly determined by the correlation of the dataset. Yet bundled curves nonetheless provided sufficient cues (width of center region, shape of lines) that participants could estimate correlation from bundled curves as readily as from polylines.

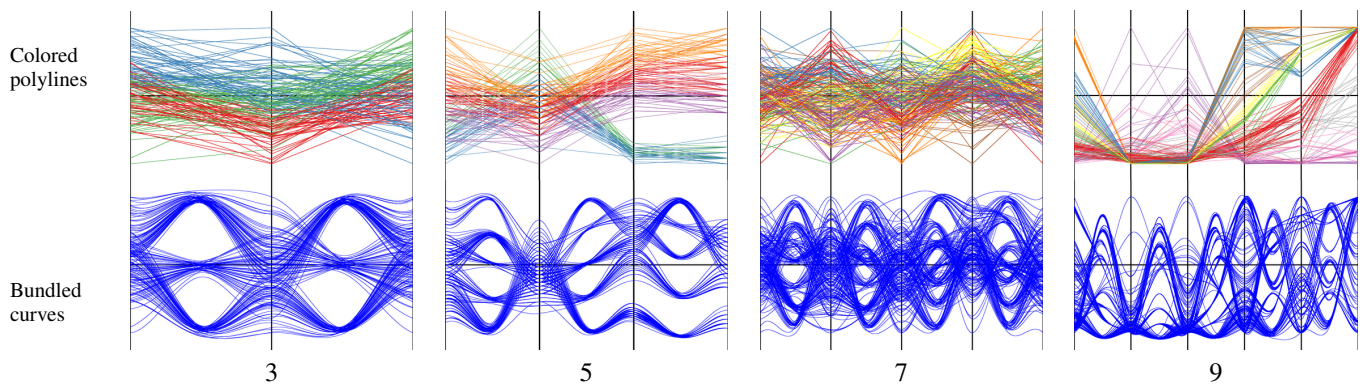

Fig. 5. Representative plots used in the cluster estimation series: polyline plots with color coding of clusters (top row), and bundled curve plots (bottom row). From left to right, the data sets are g160 (number of $k$-means clusters $k = 3$), iris ($k = 5$), g200 ($k = 7$), and netperf ($k = 9$).

The cluster counting series demonstrates that viewers could identify clusters through their bundles. This is not surprising, as bundling provides a strong cue of cluster identity. Participants likely determined cluster membership by looking at the bundle axes, where bundling has its strongest effect. In effect, a bundled curve plot uses different regions to geometrically represent the spread and the clustering of the dataset. The spread of values for a cluster is represented at the value axis. The cluster identity of a datum is represented at the bundle axis. In contrast, polylines provide no geometric representation of cluster membership, so it must be represented using a different cue, color. Whereas polylines provide only correlation information in the inter-axis regions, bundled curves use that region to display correlation, number of clusters, and cluster membership—a much more effective use of the space.

The geometric representation of cluster and distribution must be simultaneous if the analyst is to compare the distributions of the different clusters. The bundling and $C^1$ continuity of bundled curves are essential for this comparison to occur, for these features allow the viewer to be aware of both clusters and distribution simultaneously. Bundling exploits the Gestalt principle of proximity, visually grouping the lines of a cluster in the middle of the plot. $C^1$ continuity exploits the Gestalt principle of continuity to maintain this visual grouping on the value axes, where the distribution is represented. This allows the viewer to compare the distributions of different clusters. As a secondary benefit, the $C^1$ continuity allows this cluster identification to be maintained across value axes, the membership reinforced at each bundle axis.

The same bundling strength was used for both the correlation and the cluster counting series. This demonstrates that each task can be achieved without sacrificing the other.

## V. APPLICATION

The user study demonstrated that bundling is sufficient for observers to distinguish clusters. We next complement the study by presenting an example demonstrating that a bundled parallel plot can represent the influence of clusters more directly than its polyline counterpart. We assume a use case where a clustering has been found and the analyst wishes to explore an initial hypothesis of the predictive power of cluster membership.

We use the car performance data of Ramos and Donoho,[6] a sample data set for the 1983 ASA Data Exposition. Altogether the original data set includes eight fields, of which we analyse relationships amongst seven: the number of cylinders, the miles per gallon, the displacement, the horsepower, the weight, the number of seconds to complete a quarter-mile from a standing start, and the model year. A total of 406 cars are in the dataset.

A natural basis for clustering this data is the number of cylinders in each model's engine. Clustering the data this way produces five clusters. Two clusters, the seven cars with either three or five cylinders, are so small that they increased the plot complexity with no gain in explanatory power, so we deleted them from the plot. The remaining 399 cars were grouped into clusters for four, six, or eight cylinders. A reasonable initial hypothesis is that the number of cylinders has the following correlations to the other fields:

- More cylinders should reduce the miles per gallon (negative correlation).
- More cylinders should increase displacement (positive correlation).
- More cylinders should increase horsepower (positive correlation).
- More cylinders should increase weight (positive correlation).
- More cylinders should decrease the number of seconds to accelerate over a quarter-mile (negative correlation).

For model year, there was no strong *a priori* assumption for any correlation with the other fields. We chose to plot older cars at the top, as there was a mild likelihood that they would be less efficient than newer models.

Figure 1 shows the data plotted in both polyline and bundled parallel coordinates. The direction of each value axis is selected so that the values hypothesized to correlate most directly to the maximum number of cylinders are displayed at the top (value axes with expected negative correlations have their smaller values at the top). At each bundle axis, curves within a cluster are pulled to the projections of their centroids.

Comparing the between-axes region in the two plots, the polyline version (Figure 1a) features a lot of geometry that engages the eye but provides no useful information. Between every pair of value axes many lines overlap and at each axis there are many abrupt changes of direction.

By contrast, in the between-axes region of the bundled version (Figure 1b) the geometry is informative. The projection of each cluster's centroid is indicated on the bundle axes. It is possible to immediately assess both the relative order and the magnitude of the difference between the projected centroids by number of cylinders.

Comparing the two plots at the value axes, for the polyline plot it is impossible to compare the ranges of different clusters for most variables because it is too difficult to associate a line segment with its cluster. By contrast, in the bundled plot it is easy to trace an individual curve to its cluster bundle, making it easy to compare the cluster ranges on a given axis. This includes detecting outliers within clusters, such as the most powerful car in terms of horse power within the six cylinder group. From the polyline plot, the correspondence to its cluster could not be determined without color or additional axes. Consequently, a glance at a value axis is sufficient to determine the predictive power of the clustering on that variable. For the given cars, the number of cylinders strongly predicts displacement, horsepower, and weight. For consumption and acceleration however, the number of cylinders only predicts the extreme values for each cluster.

The polyline plot is slightly more informative about the relationship between model year and acceleration (simply because they were plotted as adjacent axes). Older cars have a wider range of acceleration than more recent cars, with the quickest cars all coming from the earliest years. The bundled plot obscures the relationship of model year with any other axis because the bundle points make it impossible to track the model year to any adjacent axis.

Overall, the bundled representation is clearer because it separates distinct components of the data into distinct regions of the plot: Cluster membership is displayed on the bundle axes, while the range of individual points within the cluster is displayed on the value axes. The curves allow straightforward correlation of individual values and their membership. By contrast, the polyline representation gives equal prominence to changes within clusters and changes of the clusters as a group, making it impossible to separate them visually.

This example use case assumed that the analyst had an *a priori* hypothesis of the correlations between cluster membership and other variables. We suggest that this is the most common and important case because it represents the goal of most other use cases. For example, where there is no prior hypothesis, interactive bundled plots would allow the analyst to consider alternative relationships between the clusters and other variables by varying the direction and order of axes, together with the bundling parameters $\alpha$ and $\beta$. The clear geometric separation afforded by bundling would likely allow the analyst to converge on the best explanation far more quickly than with polyline plots.

For demonstration purposes, we also provide a video of the application scenario as supplemental material. The car data analysis is conducted with a prototype tool that allows to interactively adjust the smoothing and bundling parameters. In conjunction with standard interaction techniques such as flipping and reordering axes, the video demonstrates the strengths of bundled parallel coordinate plots in an interactive environment.

### A. Bundles aid interpreting clusters

The purpose of bundling is to highlight the cluster membership of individual points. Bundling is not itself a technique for determining clusters. Rather, it is used after the analyst has already derived clusters, whether from *a priori* domain knowledge or statistical clustering methods. Bundling aids in the interpretation of clusters in several ways.

At the local level of analysis of a single axis, bundling makes it easier to see the spread of values for one cluster and to compare spreads for several clusters. In the car example above, much of the analysis of the influence of engine cylinders was done this way. Bundling also helps compare the relative order of clusters along one axis or along two adjacent axes. These local analyses become more powerful when combined with interactive selection of the direction of each axis, as shown in the demonstration video.

At the global level of analyzing clusters across all axes, bundling is only partially useful. When the relative order of the clusters remains the same across all or most axes (perhaps after flipping some axes), clustering is sufficient to differentiate the clusters without recourse to additional methods such as colour. Figure 1 demonstrates this case for the car example.

When the relative order of clusters changes from axis to axis, however, as seen in the bundled iris and netperf plots in Figure 5, bundling is insufficient to convey the global picture of where clusters fall on every axis. The zig-zags introduced by displaying the bundling points might even make it harder for the eye to track clusters. In these cases, bundling may have to be supplemented by color to convey the global flow of each cluster. Note that color by itself may not be particularly informative about the global flow of clusters, as demonstrated by the colored polyline plots in Figure 5.

Overall, we suggest that bundling is a powerful addition to the representation of clusters in parallel coordinates, sufficient to support many kinds of local analyses and some cases of global analysis, while offering powerful synergies with colour for complex global relationships.

## VI. CONCLUSION AND FUTURE WORK

Bundled curve parallel-coordinates plots are designed to alleviate some longstanding limitations of traditional polyline plots and to geometrically reveal cluster structures specified for the input data. Bundled curve plots are constructed from piecewise cubic Bézier curves with control points judiciously selected to ensure $C^1$ continuity; this can alleviate the well-known cross-over problems for polyline plots without any additional visual aid beyond curve geometry. Cluster structures in the data are emphasized by curve bundling, pulling curves belonging to the same cluster toward their cluster centroid. The greatest advantage of bundling is that a wide range of views from abstract high-level to detailed low-level representations are easily obtained by tuning the bundling strength. The cluster centroids can further be uniformly distributed to take advantage of the whole plot area and reduce line cluttering. For large datasets, line density plots for each cluster can be employed to avoid the potential over-plotting problem. As traditional parallel coordinates are a special case of our technique, many established extensions and interaction techniques such as brushing-and-linking can be applied without further enhancements.

The user study conducted in this work supports the following conclusions: Firstly, curve bundling is effective in displaying clustering information purely based on geometry. As a consequence, the color channel can be used for other attributes or, in the case of complex cluster patterns, be used in concert with color. Secondly, with a properly chosen bundling strength, bundled curve plots retain the same strength as polyline plots in revealing correlations between visualized variables. Hence one of the core aspects of analysis using parallel coordinates carries over using bundling.

In future work, we plan to compare bundled curve plots to polylines and other curve-based approaches, as well as other approaches to visualizing cluster information. The high effectiveness of curved plots compared with polyline plots was not obvious. The results of our user study might trigger further perceptual investigations of variants of parallel-coordinates plots. It could be the case that other forms of parallel-coordinates plots might be even more effective than bundled curves—not only for cluster visualization but other applications.

Qualitative improvements of this work are also possible. One problem of interest is the automatic determination of the visualization parameters, $\alpha$ and $\beta$, as well as the arrangement of cluster centroids, guided by optimization criteria. For example, such criteria may be related to a measure of line cluttering and curve smoothness. It is also interesting to consider combining color mapping and curve density plots selectively in order to further improve the high-level visualization of clusters and low-level investigation of data items within each cluster.

This research is supported in part by the NSERC Discovery Grant of Zhang and a MITACS research grant. Special thanks to Jing Li and Jarke J. van Wijk for sharing the results of their user study.[15]

## References

[1] Ware, C. (2004) *Information Visualization: Perception for Design*. Morgan Kaufmann Publishers, second edn.

[2] Inselberg, A. (1985) The plane with parallel coordinates. *The Visual Computer*, **1**, 69–92.

[3] Inselberg, A. and Dimsdale, B. (1990) Parallel coordinates: A tool for visualizing multi-dimensional geometry. Kaufman, A. (ed.), *IEEE Visualization*, pp. 361–378, IEEE Computer Society.

[4] Addinsoft, Xlstat. http://www.xlstat.com/.

[5] Cook, D. and Swayne, D. F. (2003). GGobi software: http://www.ggobi.org/.) *Interactive and Dynamic Graphics for Data Analysis: With Examples Using R and GGobi*. Springer.

[6] Ramos, E. and Donoho, D. (1983) 1983 ASA data exposition dataset. *CMU Dataset Archive*.

[7] Theisel, H. (2000) Higher order parallel coordinates. *Workshop on Vision, Modeling, and Visualization*, pp. 415–420.

[8] Graham, M. and Kennedy, J. (2003) Using curves to enhance parallel coordinate visualisations. *Information Visualization*, pp. 10–16, IEEE Computer Society.

[9] Moustafa, R. and Wegman, E. (2006) Multivariate continuous data – parallel coordinates. Unwin, A., Theus, M., and Hofmann, H. (eds.), *Graphics of Large Datasets: Visualizing a Million*, pp. 143–156, Springer.

[10] Yuan, X., Guo, P., Xiao, H., Zhou, H., and Qu, H. (2009) Scattering points in parallel coordinates. *IEEE Transactions on Visualization and Computer Graphics*, **15**, 1001–1008.

[11] Holten, D. and van Wijk, J. J. (2010) Evaluation of cluster identification performance for different PCP variants. *Computer Graphics Forum*, **29**.

[12] Holten, D. (2006) Hierarchical edge bundles: Visualization of adjacency relations in hierarchical data. *IEEE Transactions on Visualization and Computer Graphics*, **12**, 741–748.

[13] Zhou, H., Yuan, X., Qu, H., Cui, W., and Chen, B. (2008) Visual clustering in parallel coordinates. *Computer Graphics Forum*, **27**, 1047–1054.

[14] McDonnell, K. T. and Mueller, K. (2008) Illustrative parallel coordinates. *Computer Graphics Forum*, **27**, 1031–1038.

[15] Li, J., Martens, J.-B., and van Wijk, J. J. (2010) Judging correlation from scatterplots and parallel coordinate plots. *Information Visualization*, **9**, 13–30.

[16] Inselberg, A. (2009) *Parallel Coordinates: Visual Multidimensional Geometry and Its Applications*. Springer.

[17] Wegman, E. (1990) Hyperdimensional data analysis using parallel coordinates. *Journal of the American Statistical Association*, **411**, 664.

[18] Miller, J. J. and Wegman, E. J. (1991) Construction of line densities for parallel coordinate plots. Buja, A. and Tukey, P. (eds.), *Computing and Graphics in Statistics*, pp. 107–123, Springer.

[19] Wegman, E. and Luo, Q. (1997) High dimensional clustering using parallel coordinates and the grand tour. *Computing Science and Statistics*, **28**, 361–368.

[20] Rodrigues, J. F., Jr., Traina, A. J. M., and Traina, C., Jr. (2003) Frequency plot and relevance plot to enhance visual data exploration. de Oliveira, M. C. F. and Junior, R. M. C. (eds.), *Symposium on Computer Graphics and Image Processing (SIBGRAPI)*, pp. 117–124, IEEE Computer Society.

[21] Artero, A. O., de Oliveira, M. C. F., and Levkowitz, H.

(2004) Uncovering clusters in crowded parallel coordinates visualizations. Ward, M. and Munzner, T. (eds.), *IEEE Symposium on Information Visualization*, pp. 81–88, IEEE Computer Society.

[22] Johansson, J., Ljung, P., Jern, M., and Cooper, M. (2005) Revealing structure within clustered parallel coordinates displays. Stasko, J. and Ward, M. (eds.), *IEEE Symposium on Information Visualization*, pp. 125–132, IEEE Computer Society.

[23] Heinrich, J. and Weiskopf, D. (2009) Continuous Parallel Coordinates. *IEEE Transactions on Visualization and Computer Graphics*, **15**, 1531–8.

[24] Jain, A. K. and Dubes, R. C. (1988) *Algorithms for Clustering Data*. Prentice-Hall.

[25] Fua, Y.-H., Ward, M. O., and Rundensteiner, E. A. (1999) Hierarchical parallel coordinates for exploration of large datasets. David Ebert, B. H., Markus Gross (ed.), *IEEE Visualization*, pp. 43–50, IEEE Computer Society.

[26] Berthold, M. R. and Hall, L. O. (2003) Visualizing fuzzy points in parallel coordinates. *IEEE Transactions on Fuzzy Systems*, **11**, 369–374.

[27] Novotny, M. and Hauser, H. (2006) Outlier-preserving focus+context visualization in parallel coordinates. *IEEE Transactions on Visualization and Computer Graphics*, **12**, 893–900.

[28] Zhou, H., Cui, W., Qu, H., Wu, Y., Yuan, X., and Zhuo, W. (2009) Splatting the lines in parallel coordinates. *Computer Graphics Forum*, **28**, 759–766.

[29] Lanzenberger, M., Miksch, S., and Pohl, M. (2005) Exploring highly structured data: a comparative study of stardinates and parallel coordinates. *Information Visualisation*, pp. 312–320, IEEE Computer Society.

[30] Henley, M., Hagen, M., and Bergeron, R. D. (2007) Evaluating two visualization techniques for genome comparison. *International Conference on Information Visualization (IV)*, pp. 551–558.

[31] ten Caat, M. and Maurits, N. M. (2007) Design and evaluation of tiled parallel coordinate visualization of multichannel EEG data. *IEEE Transactions on Visualization and Computer Graphics*, **13**, 70–79.

[32] Johansson, J., Forsell, C., Lind, M., and Cooper, M. (2008) Perceiving patterns in parallel coordinates: determining thresholds for identification of relationships. *Information Visualization*, **7**, 152–162.

[33] Farin, G. (2001) *Curves and Surfaces for CAGD: A Practical Guide*. Morgan Kaufmann.

[34] Color vision test. http://vision.healthcommunities.com/HealthProfiler/healthpro_cb.shtml.

[35] Harrower, M. A. and Brewer, C. A. (2003) ColorBrewer.org: an online tool for selecting color schemes for maps. *The Cartographic Journal*, **40**, 27–37.

[36] Xmdvtool home page. http://davis.wpi.edu/xmdv/datasets.html.